\documentstyle[aps,pre,amssymb]{revtex}

\input epsf

\begin{document}
\title{Biased Diffusion with Correlated Noise%
\medskip }
\author{H.K. Janssen$^{1}$ and B. Schmittmann$^{2}$ \bigskip}
\address{$^{1}$Institut f\"{u}r Theoretische Physik III, \\
Heinrich-Heine-Universit\"{a}t, D-40225 D\"{u}sseldorf, Germany;\\
$^{2}$Center for Stochastic Processes in Science and Engineering and\\
Department of Physics, \\
Virginia Tech, Blacksburg, Va 24061-0435, USA.}
\date{\today}
\maketitle

\begin{abstract}
The diffusion of hard-core particles subject to a global bias is described
by a nonlinear, anisotropic generalization of the diffusion equation with
conserved, local noise. Using renormalization group techniques, we analyze
the effect of an additional noise term, with spatially long-ranged
correlations, on the long-time, long-wavelength behavior of this model.
Above an upper critical dimension $d_{LR}$, the long-ranged noise is always
relevant. In contrast, for $d<d_{LR}$, we find a ``weak noise'' regime
dominated by short-range noise. As the range of the noise correlations
increases, an intricate sequence of stability exchanges between different
fixed points of the renormalization group occurs. Both smooth and
discontinuous crossovers between the associated universality classes are
observed, reflected in the scaling exponents. We discuss the necessary
techniques in some detail since they are applicable to a much wider range of
problems.
\medskip

\end{abstract}
\bigskip


\section{Introduction}

Systems coupled to multiple energy reservoirs, sustaining net transport
currents, are prevalent in nature but fall outside the fundamentally
well-understood paradigms of equilibrium statistical mechanics. The
characterization and classification of such systems, starting from simple
microscopic models, remains a key problem of current research. Since steady
states in such systems constitute the closest relatives of equilibrium
(Gibbs) states, their study has attracted much recent attention \cite{ness},
revealing a wealth of unexpected phenomena, even in simple model systems.
One of the prime, and most elementary, examples of this kind is a system of
hard-core particles undergoing biased diffusion. With appropriate boundary
conditions, this system, also known as the simple asymmetric exclusion
process (ASEP) \cite{ASEP}, can exhibit a number of surprising features,
including anomalous diffusion \cite{ASEP,vBKS,JS}, long-range spatial
correlations \cite{ZS}, and shocks \cite{shocks}. One of its key
characteristics is the breaking of the usual fluctuation-dissipation theorem 
\cite{FDT}.

In the following, we will discuss the effect of long-range correlated noise
on this model, formulated via a Langevin equation in continuous space and
time. Generated, for example, by correlated fluctuations in the strength of
the bias, the additional operator acts in the one-dimensional subspace
(labelled{\bf \ }``longitudinal'') selected by the bias. Its momentum
dependence takes the form $q_{\Vert }{}^{2(1-\alpha )}$, with $0\leq \alpha
\leq 1$. The two extreme values, $\alpha =0$ and $\alpha =1$, are of
particular interest. The choice $\alpha =0$ corresponds to the standard
ASEP. In contrast, $\alpha =1$ leads to a Langevin equation with {\em %
non-conserved} noise, describing biased diffusion of particles which can
occasionally be created or annihilated. First introduced by Hwa and Kardar 
\cite{HK} as a continuum description for sliding avalanches in sandpile
models, it was analyzed in more detail by Becker and Janssen \cite{BJ}.

The key advantage of the correlated noise term is that it allows us to
capture the {\em crossover} between these two models, characterized,
respectively, by conserved and non-conserved noise. Thus, using field
theoretic techniques, we discuss the full renormalization group (RG) flow in 
$\left( \alpha ,d\right) $ space, for $0\leq \alpha \leq 1$. While the limit 
$\alpha \rightarrow 1$ presents no difficulties, the opposite limit $\alpha
\rightarrow 0$ is far more subtle: we will see that, just below the upper
critical dimension $d_{c}\left( \alpha \right) $, {\em two} nontrivial
renormalizations are required to render the $\alpha =0$ theory finite, in
contrast to a {\em single} one if $\alpha $ is positive and of $O(1)$,
leading to an apparent discontinuity in the critical exponents. A similar
situation arises in standard $\phi ^{4}$-theory with long-ranged
interactions \cite{HN}. There, however, the discontinuity is entirely
spurious and can be removed: letting $\varepsilon =d_{c}\left( 0\right) -d$
denote the distance from the upper critical dimension of the short-range
theory, the key is to recognize \cite{HN} that there is a region of {\em %
small} $\alpha =O\left( \varepsilon \right) $, where a careful analysis of
the RG flow reveals a smooth crossover between the $\alpha =0$ and the $%
\alpha =O\left( 1\right) $ theories. Thus, all universal scaling properties
are shown to depend continuously on $\alpha $ and $d$. Here, in contrast,
the limit $\alpha \rightarrow 0$ is even more intricate: while the
short-range theory is controlled by two fixed points, only a single one
remains in the long-range model. Thus, only one of the two short-range fixed
points is smoothly connected to the long-range fixed point, leading to a
continuous crossover. The other short-range fixed point eventually loses its
stability, accompanied by a discontinuity in critical exponents.

A second feature of our model concerns its close relationship to the KPZ 
\cite{KPZ} or Burgers' \cite{Burgers} equation with correlated noise \cite
{MHKZ,FJT}. It is well known that the continuum description for the ASEP and
the noisy Burgers equation coincide in one spatial dimension. In contrast to
the Burgers equation, the $\varepsilon $-expansion for biased diffusion
around $d_{c}=2$ presents no difficulties, and numerous results can be
obtained exactly, to all orders in perturbation theory. These may therefore
be analytically continued to $d=1$ and are expected to hold for both, the
ASEP and the KPZ equation. In fact, to the extent that exact solutions \cite
{exact} are available, this expectation is indeed confirmed. Thus, our
analysis bears some relation to the behavior of the one-dimensional Burgers
equation with correlated noise.

This paper is organized as follows. We first introduce the Langevin equation
for the ``short-range'' (SR) version of our model, corresponding to the
usual $d$-dimensional ASEP with translational invariance. We then add a
second noise term with long-range spatial correlations and propose two
possible microscopic mechanisms for such a term. Turning to the RG analysis,
we first summarize the SR case, $\alpha =0$. The ``long-range''\ (LR) theory
associated with $\alpha =O\left( 1\right) $ is presented next, and the limit 
$\alpha \rightarrow 1$ is discussed. Finally, we introduce the ``hybrid''
model, characterized by $\alpha =O\left( \varepsilon \right) $, which
interpolates between the SR ($\alpha =0$) and LR ($\alpha =O\left( 1\right) $%
) theories. Its RG flow is computed in a double expansion, where both $%
\alpha $ and $\varepsilon $ are small parameters. The different fixed points
are interpreted and their stability is evaluated, illustrating how the
subtle crossover from $\alpha =0$ to $\alpha =O\left( 1\right) $ occurs. We
conclude with some comments and open questions.

\section{The Model}

We consider a $d$-dimensional system of hard-core particles, which are
allowed to diffuse on a regular lattice with fully periodic boundary
conditions. The bias, reminiscent of an electric field ${\bf E}=E{\bf e}%
_{\Vert }$ acting on charges, favors particle moves along a specific
(``longitudinal'') direction along unit vector ${\bf e}_{\Vert }$. The
long-time, long-wavelength properties of this model are most conveniently
captured by a coarse-grained description in continuous space and time. Here,
only a single slow variable is needed, namely, the local particle density $%
c\left( {\bf r},t\right) $. Since the associated Langevin equation has been
discussed previously \cite{vBKS,JS}, we describe it only briefly here. Due
to particle number conservation, it takes the form of a continuity equation, 
$\partial _{t}c+\nabla {\bf j}=0$. In addition to a diffusive term, the
current ${\bf j}$ contains an Ohmic contribution, ${\bf j}_{E}=\kappa \left(
c\right) {\bf E}$, induced by the bias, and a Gaussian noise term, modelling
the fast degrees of freedom. By virtue of the excluded volume constraint,
the conductivity $\kappa \left( c\right) $ must vanish with both particle
and hole density, i.e., $\kappa \left( c\right) \propto c\left( 1-c\right) $%
. Aiming for a perturbation expansion around a field with zero average, we
introduce the deviations $s\left( {\bf r},t\right) =c\left( {\bf r},t\right)
-\bar{c}$ from the average density $\bar{c}$. A term linear in $s$, arising
from the expansion of $\kappa \left( c\right) $, can be absorbed via a
Galilei transformation. Finally, we allow for anisotropies in the diffusion
coefficient and the correlations of the Langevin noise, since the bias
singles out a preferred direction. Neglecting contributions that are
irrelevant in the long-wavelength, long-time limit, the Langevin equation
can be summarized in the form \cite{JS}: 
\begin{equation}
\partial _{t}s=\lambda \left\{ \left( \nabla _{\bot }^{2}+\rho \partial
^{2}\right) s+\frac{g}{2}\partial s^{2}\right\} +\eta \left( \vec{r}%
,t\right) .  \label{LE}
\end{equation}
Here, $\nabla _{\bot }$ ($\partial $) denote the gradients in the transverse
(longitudinal) subspace, and the kinetic coefficient $\lambda $ defines a
time scale. The coupling $g\propto E$ captures the effects of the drive. The
noise $\eta \left( {\bf r},t\right) $ is Gaussian so that two moments
suffice to characterize its full distribution (after a suitable rescaling of
the variables):

\begin{eqnarray}
\left\langle \eta \left( {\bf r},t\right) \right\rangle &=&0,  \nonumber \\
\left\langle \eta \left( {\bf r},t\right) \eta \left( {\bf r}^{\prime
},t^{\prime }\right) \right\rangle &=&-2\lambda \left( \nabla _{\bot
}^{2}+\sigma \partial ^{2}\right) \delta \left( {\bf r}-{\bf r}^{\prime
}\right) \delta \left( t-t^{\prime }\right) .  \label{SRN}
\end{eqnarray}
The differential operator $\left( \nabla _{\bot }^{2}+\sigma \partial
^{2}\right) $ in the second moment ensures that the conservation law is
strictly obeyed. Clearly, the correlations described by this form are purely
local, so that we will refer to Eqns.~(\ref{LE},\ref{SRN}) as the
``short-range'' (SR) theory. Its universal properties have been discussed in 
\cite{JS,ZS}.

The simplest way of introducing a correlated noise into this equation is to 
{\em add} a long-range correlated noise to the local one. Thus, Eqn.~(\ref
{SRN}) is amended to 
\begin{eqnarray}
\left\langle \eta \left( {\bf r},t\right) \right\rangle &=&0,  \nonumber \\
\left\langle \eta \left( {\bf r},t\right) \eta \left( {\bf r},t^{\prime
}\right) \right\rangle &=&2\lambda \left[ -\left( \nabla _{\bot }^{2}+\sigma
\partial ^{2}\right) +b\left( -\partial ^{2}\right) ^{1-\alpha }\right]
\delta \left( {\bf r}-{\bf r}^{\prime }\right) \delta \left( t-t^{\prime
}\right) .  \label{LRN}
\end{eqnarray}
The new operator $b\left( -\partial ^{2}\right) ^{1-\alpha }$ reflects a
power-law decay of the noise correlations in real space, giving rise to a $%
q_{\Vert }^{2\left( 1-\alpha \right) }$ momentum dependence in Fourier
space. Clearly, setting $\alpha $ to zero reproduces the SR theory, albeit
with $\sigma $ replaced by $\sigma +b$. In contrast, choosing $\alpha =1$
generates a non-conserved noise, playing the role of a random source term in
Eqn.~(\ref{LE}). Note that the vanishing of the first moment ensures that
the particle density, while not strictly invariant under the dynamics,
remains conserved {\em on average}. The long-time, long-wavelength behavior
of this theory has been analyzed in \cite{HK,BJ}. Other choices of
long-ranged noise correlations are of course possible \cite{JSlong}. The
version adopted above is ``minimal'' in the sense that, first, it leads to a
nontrivial, renormalizable theory, and second, it allows us to discuss the
full crossover between the conserved ($\alpha =0$) and non-conserved ($%
\alpha =1$) cases using just one additional operator. Microscopically, such
long-ranged contributions to the Langevin noise can be generated in two
ways: one option is to impose a bias which consists of a spatially uniform
and a locally random component, with the latter controlled by a distribution
with long-range correlations. Alternatively, one can retain the strictly
uniform bias but randomly add or remove particles, according to a
long-ranged distribution. More details will be presented in \cite{JSlong}.

Due to the presence of the nonlinearity, the computation of averages over
the noise distribution requires a perturbative approach, coupled with
renormalization group methods. For these purposes, it is most convenient to
recast the Langevin equations as dynamic functionals \cite{dynfun}.
Introducing a Martin-Siggia-Rose \cite{MSR} response field $\tilde{s}\left( 
{\bf r},t\right) $, we can recast Eqns.~(\ref{LE},\ref{LRN}) in the form 
\begin{eqnarray}
{\cal J}\left[ \tilde{s},s\right] &=&\int dt\int d^{d}r\left\{ \tilde{s}%
\partial _{t}s-\lambda \tilde{s}\left[ \left( \nabla _{\bot }^{2}+\rho
\partial ^{2}\right) s+\frac{g}{2}\partial s^{2}\right] \right.  \nonumber \\
&&\left. +\lambda \tilde{s}\left[ \left( \nabla _{\bot }^{2}+\sigma \partial
^{2}\right) -b\left( -\partial ^{2}\right) ^{1-\alpha }\right] \tilde{s}%
\right\}  \label{DF}
\end{eqnarray}
so that correlation and response functions can be computed as functional
averages with weight $\exp \left( -{\cal J}\right) $. We stress that the SR
model is recovered from this general expression, simply by setting $b=0$.

As a first step towards the RG analysis of this model, we discuss its
scaling symmetries. First, under a global rescaling of all coordinates, $%
\vec{r}\rightarrow \mu ^{-1}\vec{r}$, we find a scale invariant theory{\bf \ 
}(as long as we neglect any cutoff-dependence) provided $\lambda
t\rightarrow \mu ^{-2}\lambda t$, $s\rightarrow \mu ^{\left( d-2\alpha
\right) /2}s$ and $\tilde{s}\rightarrow \mu ^{\left( d+2\alpha \right) /2}%
\tilde{s}${\bf .} Moreover, we obtain $g\rightarrow \mu ^{\left( 1+\alpha
\right) -d/2}g$ which allows us to identify the upper critical dimension $%
d_{c}\left( \alpha \right) =2\left( 1+\alpha \right) $. To avoid confusion,
we will reserve the symbol $\varepsilon $ for the SR case, i.e., $%
\varepsilon \equiv 2-d$. For the general case, we define $\bar{\varepsilon}%
\equiv d_{c}\left( \alpha \right) -d=2\left( 1+\alpha \right) -d$. Second,
we may rescale the parallel coordinate alone, due to the anisotropy of the
model: under $x_{\Vert }\rightarrow \beta x_{\Vert }$, scale invariance is
retained if $s\rightarrow \beta ^{-1/2}s$, $\tilde{s}\rightarrow \beta
^{-1/2}s$, $\rho \rightarrow \beta ^{2}\rho $, $\sigma \rightarrow \beta
^{2}\sigma $, $b\rightarrow \beta ^{2\left( 1-\alpha \right) }b$ and $%
g\rightarrow \beta ^{3/2}g$. This permits us to identify a set of {\em %
effective} couplings, namely, $u=G_{d}g^{2}\rho ^{-5/2}\sigma $, $w=\rho
/\sigma $ and $v=b\rho ^{\alpha }/\sigma $. These will emerge quite
naturally in perturbation theory, combined with appropriate powers of $\mu $
which absorb the remaining momentum dependence. For convenience, a geometric
factor $G_{d}=S_{d}/\left( 2\pi \right) ^{d}$, with $S_{d}$ the surface of
the $d$-dimensional unit sphere, has been absorbed into $u$. For later
reference, we note that only the positive octant $u\geq 0,w\geq 0$ and $%
v\geq 0$ corresponds to physical theories even though the RG flow can be
discussed in a larger parameter space. Finally, the dynamic functional (\ref
{DF}) exhibits a nontrivial continuous symmetry, with parameter $a$, which
is characterized by the ``Galilean'' transformation $s\left( {\bf r}%
,t\right) \rightarrow s\left( {\bf r}+\lambda ag{\bf e}_{\Vert }t,t\right)
+a $, $\tilde{s}\left( {\bf r},t\right) \rightarrow \tilde{s}\left( {\bf r}%
+\lambda ag{\bf e}_{\Vert }t,t\right) $.

So far, we were able to consider the case of general $0\leq \alpha \leq 1$.
The next section will show, however, that this procedure cannot be continued
for the actual perturbative calculation of correlation and response
functions.

\section{Renormalization Group Analysis}

We now turn to perturbation theory, with the goal of identifying those
correlation and response functions that require renormalization, due to the
presence of ultraviolet (UV) divergences in Feynman diagrams. As usual, we
focus on the one-particle irreducible (1PI) vertex functions with $\tilde{n}$
($n$) external $\tilde{s}$- ($s$-) legs, $\Gamma _{\tilde{n}n}$. These will
be computed in dimensional regularization using minimal subtraction. Since
the models associated with $\alpha =0$ and $\alpha =1$ have been discussed
previously (in \cite{JS,ZS} and \cite{HK,BJ}, respectively), we only briefly
summarize their RG structure, focusing predominantly on the theory with $%
0<\alpha <1$.

\subsection{The short-range theory: $\alpha =0$.}

We first review the model with $\alpha =0$ \cite{JS,ZS}. Here, the coupling $%
b$ and the effective $v$ are redundant and will be set to zero. The upper
critical dimension is $d_{c}\left( 0\right) =2$. Straightforward dimensional
analysis shows that there are three naively divergent vertex functions: $%
\Gamma _{11}$, $\Gamma _{20}$ and $\Gamma _{12}$. A Ward identity based on
the ``Galilean'' symmetry shows, however, that $\Gamma _{12}$ requires no
additional renormalization once $\Gamma _{11}$ and $\Gamma _{20}$ have been
rendered finite. Letting the superscript $\mathaccent"7017{}\ $denote bare
couplings, the renormalized couplings are defined in the usual way: $%
\mathaccent"7017\rho =Z_{\rho }^{SR}\rho $, $\mathaccent"7017\sigma =Z_{\rho
}^{SR}\sigma $, $\mathaccent"7017g=Z_{g}g\mu ^{\varepsilon /2}$, where $%
\varepsilon =2-d$. The parameter $\mu $ sets a typical momentum scale which
will control the RG flow. All SR functions are marked with an explicit
superscript to distinguish them from their finite $\alpha $ counterparts.
The UV divergences in $\Gamma _{11}$ and $\Gamma _{20}$ give rise to
nontrivial $Z$-factors for $\rho $ and $\sigma $, which can be computed
perturbatively in $u$. A one-loop calculation yields $Z_{\rho }^{SR}=u\left(
3w+1\right) /\left( 16\varepsilon \right) +O\left( u^{2}\right) $ and $%
Z_{\sigma }^{SR}=u\left( 3w^{2}+2w+3\right) /\left( 32\varepsilon \right)
+O\left( u^{2}\right) $. We emphasize that $w$ is {\em not} treated
perturbatively here; therefore, $O\left( 1\right) $ fixed points for $w$
should not come as a surprise later. Finally, $Z_{g}=1$ to all orders in $u$%
, due to the ``Galilean'' invariance. To complete the discussion, we
introduce the Wilson functions 
\begin{eqnarray}
\zeta _{\rho }^{SR} &\equiv &\left. \mu \partial _{\mu }\ln \rho \right|
_{bare}=-u\frac{3w+1}{16}+O\left( u^{2}\right) ,  \nonumber \\
\zeta _{\sigma }^{SR} &\equiv &\left. \mu \partial _{\mu }\ln \sigma \right|
_{bare}=-u\frac{3w^{2}+2w+3}{32}+O\left( u^{2}\right) .  \label{SRWF}
\end{eqnarray}
The RG flow is expressed through the $\mu $-dependence of the renormalized
couplings $u=G_{d}g^{2}\rho ^{-5/2}\sigma $ and $w=\rho /\sigma $,
especially in the scaling limit $\mu \rightarrow 0$: 
\begin{eqnarray}
\beta _{u}^{SR}\left( u,w\right) &\equiv &\left. \mu \partial _{\mu
}u\right| _{bare}=-\left[ \varepsilon +\frac{5}{2}\zeta _{\rho }^{SR}-\zeta
_{\sigma }^{SR}\right] u,  \nonumber \\
\beta _{w}^{SR}\left( u,w\right) &\equiv &\left. \mu \partial _{\mu
}w\right| _{bare}=\Biggl[\zeta _{\rho }^{SR}-\zeta _{\sigma }^{SR}\Biggr]w.
\label{SRBF}
\end{eqnarray}
In this form, the right hand sides of Eqn.~(\ref{SRBF}) are exact to all
orders.

These flow equations possess three fixed points, marked by the vanishing of
both $\beta _{u}^{SR}$ and $\beta _{w}^{SR}$: (i) $u^{\ast }=8\varepsilon
/3+O\left( \varepsilon ^{2}\right) $, $w^{\ast }=1$; (ii) $u^{\ast
}=16\varepsilon +O\left( \varepsilon ^{2}\right) $, $w^{\ast }=0$; (iii) $%
u^{\ast }=16\varepsilon /3+O\left( \varepsilon ^{2}\right) $, $w^{\ast
}=1/3+O\left( \varepsilon \right) $; and finally a fixed line (iv) $u^{\ast
}=0$, with arbitrary $w$. The stability of these fixed points (and line) can
be expressed through the $2\times 2$ matrix ${\Bbb M}_{SR}\equiv \left(
\partial _{i}\beta _{j}^{SR}\right) ^{\ast }$, $i,j=u,v$, of derivatives of
the $\beta $-functions, evaluated at the fixed points. We find that the
fixed line (iv) is stable only for $\varepsilon <0$, i.e., $d>2$, and will
therefore be labelled a ``Gaussian'' line. Focusing on $\varepsilon >0$, we
find that (i) and (ii) are both locally stable. Fixed point (iii) is
hyperbolic and sits on the separatrix $w=1/3+O\left( \varepsilon \right) $
which separates the domains of attraction of (i) and (ii). Flow lines
starting in the region $w>1/3+O\left( \varepsilon \right) $ are attracted
towards fixed point (i), and vice versa. We refer to fixed point (i) as the
FDT-restoring short-range fixed point, since theories with $w=1$ can be
shown to possess a higher degree of symmetry, associated with satisfying the
FDT \cite{JS}. Here, Eqn.~(\ref{SRBF}) allows us to compute the fixed point
values of $\zeta _{\rho }^{SR}$ and $\zeta _{\sigma }^{SR}$ exactly, to all
orders in $\varepsilon $: $\zeta _{\rho }^{SR\ast }=\zeta _{\sigma }^{SR\ast
}=-2\varepsilon /3$. Conversely, fixed point (ii) violates the FDT, and $%
\zeta _{\rho }^{SR\ast }=-\varepsilon +O\left( \varepsilon ^{2}\right) $; $%
\zeta _{\sigma }^{SR\ast }=-3\varepsilon /2+O\left( \varepsilon ^{2}\right) $
are known only perturbatively.

Finally, we comment briefly on the scaling properties associated with these
fixed points \cite{JS,ZS}, which arise from the RG equations for the vertex
functions. For example, the dynamic density correlation function (structure
factor) $S\left( {\bf q},\omega \right) \equiv \left\langle s\left( -{\bf q}%
,-\omega \right) s\left( {\bf q},\omega \right) \right\rangle $ scales as 
\begin{equation}
S\left( {\bf q},\omega \right) =l^{-2}S\left( q_{\Vert }l^{-1-\Delta
_{SR}},q_{\perp }l^{-1},\omega l^{-2}\right) .  \label{SRS}
\end{equation}
This reflects the emergence of anomalous diffusion, characterized, near
fixed point (i), by a strong anisotropy exponent \cite{ness} $\Delta
_{SR}\equiv -\zeta _{\rho }^{SR*}/2=\left( 2-d\right) /3$, and, near fixed
point (ii) by an exponent $\Delta _{SR}\equiv -\zeta _{\rho
}^{SR*}/2=\varepsilon /2+O\left( \varepsilon ^{2}\right) $. Next, we turn to
the case of finite $\alpha $.

\subsection{The long-range models with $\alpha =O\left( 1\right) $.}

As indicated in the introduction, we will have to distinguish between values
of $\alpha $ which are $O\left( 2-d\right) $ versus those which are $O\left(
1\right) $. Here, we analyze the second case, which corresponds to true
long-range (LR) theories. The first case, being a ``hybrid'' between
short-range and long-range models, will be deferred to the next subsection.

Recalling our discussion in Section 2, the critical dimension is now $%
d_{c}\left( \alpha \right) =2\left( 1+\alpha \right) $, and we define $\bar{%
\varepsilon}=2\left( 1+\alpha \right) -d$, to be distinguished from $%
\varepsilon =2-d$. The operator $\tilde{s}\left( \nabla _{\bot }^{2}+\sigma
\partial ^{2}\right) \tilde{s}$ is clearly irrelevant compared to $b\tilde{s}%
\left( -\partial ^{2}\right) ^{1-\alpha }\tilde{s}$ and may be dropped. The
naive dimensions of the fields $s$ and $\tilde{s}$ are $\alpha $-dependent,
and as a result only {\em two} vertex functions, namely $\Gamma _{11}$ and $%
\Gamma _{12}$, are naively divergent. In contrast to the short-range case, $%
\Gamma _{20}$ is naively convergent here. Moreover, all divergent
contributions to any vertex function are polynomial in the momenta \cite{HN}%
, so that $b$ is not renormalized. The ``Galilean'' invariance still holds,
so that $\Gamma _{12}$, and hence $g$, requires no new $Z$-factor. Thus,
defining renormalized couplings according to $\mathaccent"7017\rho =Z_{\rho
}^{LR}$, $\mathaccent"7017b=Z_{b}b$, $\mathaccent"7017g=Z_{g}g\mu ^{\bar{%
\varepsilon}/2}$, we find $Z_{b}=Z_{g}=1$ to all orders, and $Z_{\rho
}^{LR}=1-\bar{u}A\left( \alpha \right) /\bar{\varepsilon}+O(\bar{u}^{2})$,
to first order in $\bar{u}=uv$. The latter is the appropriate effective
coupling here, since $\sigma $ no longer appears in the theory. The
coefficient $A\left( \alpha \right) =\Gamma \left( 1+\alpha \right) \Gamma
\left( 3/2-\alpha \right) \left( 1+2\alpha \right) /\left( 8\sqrt{\pi }%
\right) $ controls the one-loop divergence in $\Gamma _{11}$. The Wilson
functions are easily obtained: 
\begin{equation}
\zeta _{\rho }^{LR}\equiv \left. \mu \partial _{\mu }\ln \rho \right|
_{bare}=-\bar{u}A\left( \alpha \right) +O\left( \bar{u}^{2}\right)
\label{LRWF}
\end{equation}
and 
\begin{equation}
\beta _{\bar{u}}^{LR}\equiv \left. \mu \partial _{\mu }\bar{u}\right|
_{bare}=-\left[ \bar{\varepsilon}+\left( \frac{5}{2}-\alpha \right) \zeta
_{\rho }^{LR}\right] \bar{u}.  \label{LRBF}
\end{equation}
Clearly, the LR theory exhibits one (Gaussian) fixed point $\bar{u}^{*}=0$
which is easily shown to be stable above the upper critical dimension, and a
nontrivial fixed point $\bar{u}^{*}=2\bar{\varepsilon}/\left[ \left(
5-2\alpha \right) A\left( \alpha \right) \right] +O\left( \bar{\varepsilon}%
^{2}\right) $, stable below $d_{c}$. At this fixed point, the value of $%
\zeta _{\rho }^{LR}$ is again known to all orders in $\bar{\varepsilon}$,
namely, $\zeta _{\rho }^{LR*}=-2\bar{\varepsilon}/\left( 5-2\alpha \right) $.

The scaling properties of the long-range models are again easily derived
from an RG equation for the vertex functions. For comparison with the SR
case, we quote the result for the dynamic structure factor: 
\begin{equation}
S\left( {\bf q},\omega \right) =l^{-2\left( 1+\alpha \right) }S\left(
q_{\Vert }l^{-1-\Delta _{LR}},q_{\perp }l^{-1},\omega l^{-2}\right) .
\label{LRS}
\end{equation}
Once again, we observe anomalous diffusion; here, however, it is controlled
by the LR exponent $\Delta _{LR}\equiv -\zeta _{\rho }^{LR*}/2=\left(
7-d\right) /\left( 5-2\alpha \right) -1$.

It is obvious at this point that the limit $\alpha \rightarrow 0$, taken in
the Wilson function (\ref{LRWF}) or the correlation function (\ref{LRS}),
will not restore the SR theory. First, $\lim_{\alpha \rightarrow 0}\zeta
_{\rho }^{LR}=-\bar{u}/16+O\left( \bar{u}^{2}\right) $ which reproduces only
the $w=0$ component of $\zeta _{\rho }^{SR}$. Second, this naive limit
cannot generate a non-vanishing $\zeta _{\sigma }^{SR}$ which plays a key
role in the SR theory. Thus, not surprisingly, we do not observe a smooth
crossover from the SR to the LR theory, by simply letting $\alpha $ tend to
zero in this naive fashion. An elegant way of dealing with this difficulty
was suggested in \cite{HN}: By treating $\alpha $ as a small parameter of $%
O\left( \varepsilon \right) $ and expanding in both $\varepsilon $ and $%
\alpha $, one can ``magnify'' the small $\alpha $-section of $\left(
d,\alpha \right) $ space and resolve the crossover between the SR and LR
fixed points. Concerning the scaling form of the structure factor, the
prefactor $l^{-2\left( 1+\alpha \right) }$ originates entirely in the bare
dimensions of the fields. Since these are well defined under a naive $\alpha
\rightarrow 0$ limit (cf. Section 2) and remain unrenormalized, we have $%
\lim_{\alpha \rightarrow 0}l^{-2\left( 1+\alpha \right) }=l^{-2}$ without
further complications. The exponent $\Delta _{LR}$, however, must be tracked
much more carefully.

Before turning to these subtleties, we briefly consider the other limit, $%
\alpha \rightarrow 1$. First, we summarize the $\alpha =1$ theory \cite{BJ},
with $\bar{\varepsilon}=4-d$, in its own right. Here, $b$ measures the
strength of the (non-conserved) noise. The effective coupling is $\bar{u}%
=G_{d}g^{2}\rho ^{-3/2}b$. According to \cite{BJ}, there is only one
nontrivial $Z$-factor, namely $Z_{\rho }=1-3\bar{u}/\left( 8\varepsilon
\right) +O\left( \bar{u}^{2}\right) $. Again, $Z_{b}=Z_{u}=1$ to all orders.
Thus, the theories with $\alpha =1$ and $0<\alpha <1$ exhibit the same set
of UV divergences. From the preceding discussion, it is therefore clear that
we may anticipate a smooth crossover here. This is explicitly confirmed by
the results of \cite{BJ} for $\alpha =1$: two fixed points $\bar{u}^{*}=0$
and $\bar{u}^{*}=16\varepsilon /9+O\left( \varepsilon ^{2}\right) $ are
found, stable above and below $d_{c}=4$, respectively. At the nontrivial
fixed point, $\Delta =-\zeta _{\rho }^{*}/2$ takes the exact value $%
\varepsilon /3$. It is easy to check that this agrees with the $\alpha
\rightarrow 1$ limit of Eqns.~(\ref{LRWF}-\ref{LRS}), so that this limit is
continuous.

\subsection{The hybrid theory: $\alpha =O\left( \varepsilon \right) $.}

Finally, we turn to the analysis of the key region in $\left( d,\alpha
\right) $ space, namely, $\alpha =O\left( \varepsilon \right) $. The naive $%
\alpha \rightarrow 0$ limit presupposes $\varepsilon \ll \alpha $, and hence
fails to resolve the crossover between the SR and LR theories which occurs
for $\alpha =O\left( \varepsilon \right) $. With both $\alpha $ and $%
\varepsilon $ small, we follow \cite{HN} and analyze the general dynamic
functional, Eqn.~(\ref{DF}), near the upper critical dimension of the SR
theory, i.e., $d=2$. Both couplings $\sigma $ and $b$ will be retained, and
their RG flow will be studied. We refer to this model as the ``hybrid''
theory, since it contains the vital elements of {\em both} SR and LR cases.
Thus, the hybrid theory will exhibit a well-defined $\alpha \rightarrow 0$
limit. Once again, we will be able to obtain a number of results to all
orders in $\varepsilon =2-d$.

Our first task will be to identify a set of suitable couplings which allow
us to interpolate between the SR and the LR theories. Guided by \cite{HN},
we consider the general structure of the Wilson functions in the SR and the
hybrid model. The appropriate effective couplings will be those that map
those functions onto one another.

To obtain a general form for the Wilson functions, we begin with the $Z$%
-factors and construct them order by order in perturbation theory. Focusing
on the SR theory first, a typical graph of $\Gamma _{11}$, at $L$-loop,
contains $2L$ vertices, $L$ (bare) correlators and $2L-1$ (bare)
propagators. Each vertex carries a factor of $\mathaccent"7017gq_{\Vert }$,
and each correlator contributes a factor $\left( q_{\perp }^{2}+\sigma
q_{\Vert }^{2}\right) $, where $\left( \omega ;{\bf q}\right) =\left( \omega
;q_{\Vert },{\bf q}_{\perp }\right) $ denote the frequency and momentum of
the corresponding line. Symmetry requires that $2$, out of the $L$ parallel
momentum factors contributed by the vertices, will not be integrated over.
Moreover, all denominators are generically of the form $\left( i\omega
\lambda ^{-1}+q_{\perp }^{2}+\rho q_{\Vert }^{2}\right) $. Changing the
parallel integration variable from $q_{\Vert }$ to $\rho ^{1/2}q_{\Vert }$,
it is straightforward to show that such a graph will give rise to a
contribution of the form $u^{L}\sum_{m=0}^{L}w^{m}A_{Lm}/\left( L\varepsilon
\right) +O\left( \varepsilon ^{-2}\right) $ in $Z_{\rho }^{SR}$. Here, the $%
A_{Lm}$ are the numerical coefficients of the $O\left( \varepsilon
^{-1}\right) $ poles, in minimal subtraction. Higher order poles appear
also, of course, but cancel in the Wilson functions and need not be
considered for this reason. Generic $L$-loop graphs of $\Gamma _{20}$ have a
similar structure, with $L+1$ correlators and $2L-2$ propagators, and the
coefficients of the lowest order $\varepsilon $-poles will be denoted by $%
B_{Lm}$ here. Collecting, we obtain the $Z$-factors in the general form: 
\begin{eqnarray}
Z_{\rho }^{SR} &=&1+\sum_{L=1}^{\infty }u^{L}\sum_{m=0}^{L}w^{m}\frac{A_{Lm}%
}{L\varepsilon }+O\left( \varepsilon ^{-2}\right) ,  \nonumber \\
Z_{\sigma }^{SR} &=&1+\sum_{L=1}^{\infty }u^{L}\sum_{m=0}^{L+1}w^{m}\frac{%
B_{Lm}}{L\varepsilon }+O\left( \varepsilon ^{-2}\right) .  \label{SRZ}
\end{eqnarray}
The general form of the Wilson functions is easily found from the above: 
\begin{eqnarray}
\zeta _{\rho }^{SR} &=&\sum_{L=1}^{\infty }u^{L}\sum_{m=0}^{L}w^{m}A_{Lm}, 
\nonumber \\
\zeta _{\sigma }^{SR} &=&\sum_{L=1}^{\infty
}u^{L}\sum_{m=0}^{L+1}w^{m}B_{Lm}.  \label{SRW}
\end{eqnarray}

Next, we consider the hybrid case. Two key differences emerge: First, each
correlator now contributes a factor of $\left( q_{\perp }^{2}+\sigma
q_{\Vert }^{2}+bq_{\Vert }^{2(1-\alpha )}\right) $ to a given diagram. This
product gives rise to a sum of individual terms each of which contains $L-n$
factors of type $\left( q_{\perp }^{2}+\sigma q_{\Vert }^{2}\right) $ and $n$
factors of type $bq_{\Vert }^{2(1-\alpha )}$, with $n=0,1,...,L$. Second,
since $\alpha $ is now of $O\left( \varepsilon \right) $, the poles in these
individual contributions are proportional to $1/\left( L\varepsilon
+2n\alpha \right) $ \cite{HN}. Following the same reasoning as above, and
recalling the effective coupling $v=b\rho ^{\alpha }/\sigma $, one finds
quite readily that 
\begin{eqnarray}
Z_{\rho } &=&1+\sum_{L=1}^{\infty }u^{L}\sum_{n=0}^{L}\sum_{m=0}^{L-n}%
{L \choose n}%
w^{m}v^{n}\frac{A_{Lmn}}{L\varepsilon +2n\alpha }+O\left( \varepsilon
^{-2}\right) ,  \nonumber \\
Z_{\sigma } &=&1+\sum_{L=1}^{\infty }u^{L}\sum_{n=0}^{L+1}\sum_{m=0}^{L+1-n}%
{L+1 \choose n}%
w^{m}v^{n}\frac{B_{Lmn}}{L\varepsilon +2n\alpha }+O\left( \varepsilon
^{-2}\right) .
\end{eqnarray}
Here, $A_{Lmn}$ and $B_{Lmn}$ are the numerical coefficients of each pole.
In minimal subtraction, they are independent of both $\varepsilon $ and $%
\alpha $, since the latter is also $O(\varepsilon )$ here. Clearly, the
associated Wilson functions take the form 
\begin{eqnarray}
\zeta _{\rho } &=&\sum_{L=1}^{\infty }u^{L}\sum_{n=0}^{L}\sum_{m=0}^{L-n}%
{L \choose n}%
w^{m}v^{n}A_{Lmn},  \nonumber \\
\zeta _{\sigma } &=&\sum_{L=1}^{\infty
}u^{L}\sum_{n=0}^{L+1}\sum_{m=0}^{L+1-n}%
{L+1 \choose n}%
w^{m}v^{n}B_{Lmn}.  \label{LRW}
\end{eqnarray}
Next, we establish a relationship between the coefficients $\left(
A_{Lm},B_{Lm}\right) $ and $\left( A_{Lmn},B_{Lmn}\right) $, by considering
the $\alpha \rightarrow 0$ limit of the hybrid theory. In this limit, the
dynamic functional, Eqn.~(\ref{DF}), simply reduces to that of the SR
theory, with $\sigma $ replaced by $\sigma +b$. Therefore, we can relate $%
\lim_{\alpha \rightarrow 0}Z_{\rho }$ and $\lim_{\alpha \rightarrow
0}Z_{\sigma }$ to $Z_{\rho }^{SR}$ and $Z_{\sigma }^{SR}$. Due to the
replacement of $\sigma $ by $\sigma +b$, the SR $Z$-factors depend on a
modified set of couplings. Defining 
\[
\bar{u}\equiv u\left( 1+v\right) ,\hspace{0.25in}\bar{w}\equiv \frac{w}{1+v}%
, 
\]
we obtain: 
\begin{eqnarray}
\lim_{\alpha \rightarrow 0}Z_{\rho }\left( u,w,v\right) &=&Z_{\rho
}^{SR}\left( \bar{u},\bar{w}\right) ,  \nonumber \\
\lim_{\alpha \rightarrow 0}Z_{\sigma }\left( u,w,v\right) &=&\left(
1+v\right) Z_{\sigma }^{SR}\left( \bar{u},\bar{w}\right) -v.  \label{limZ}
\end{eqnarray}
Inserting the explicit forms for the $Z$-factors into Eqn.~(\ref{limZ}) and
recalling that $\left( A_{Lmn},B_{Lmn}\right) $ are independent of $\alpha $%
, we read off the relation between the two sets of coefficients: 
\begin{eqnarray}
{L \choose n}%
A_{Lmn} &=&%
{L-m \choose n}%
A_{Lm},  \nonumber \\
{%
{L+1 \choose n}%
}B_{Lmn} &=&%
{L+1-m \choose n}%
B_{Lm}.  \label{AB}
\end{eqnarray}
Combining Eqns.~(\ref{AB}) and (\ref{LRW}) provides us with an{\em \ identity%
} between SR and hybrid Wilson functions: 
\begin{eqnarray}
\zeta _{\rho }(u,w,v) &=&\zeta _{\rho }^{SR}(\bar{u},\bar{w}),  \nonumber \\
\zeta _{\sigma }(u,w,v) &=&(1+v)\zeta _{\sigma }^{SR}(\bar{u},\bar{w}).
\label{SRH}
\end{eqnarray}
Here, $\zeta _{\rho }^{SR}$ and $\zeta _{\sigma }^{SR}$ denote the Wilson
functions of the SR model, Eqn.~(\ref{SRWF}), evaluated at $\left( \bar{u},%
\bar{w}\right) $. In the following, we use the abbreviated notation $\bar{%
\zeta}_{\rho }\equiv \zeta _{\rho }^{SR}(\bar{u},\bar{w})$ and $\bar{\zeta}%
_{\sigma }\equiv \zeta _{\sigma }^{SR}(\bar{u},\bar{w})$.

Eqns.~(\ref{SRH}) form the basis for the remainder of the paper: they hold
the key for the discussion of the hybrid theory and for the desired matching
between LR and SR models. It is particularly gratifying that they are valid%
{\em \ to all orders} in perturbation theory. An explicit calculation of the
hybrid Wilson functions, instead of the general considerations presented
above, would establish Eqns.~(\ref{SRH}) only up to a given order.

To discuss the RG flow for the hybrid theory, we compute the $\beta $%
-functions for the couplings $(\bar{u},\bar{w},v)$: 
\begin{eqnarray}
\beta _{\bar{u}}\left( \bar{u},\bar{w},v\right) &=&\left. \mu \partial _{\mu
}\bar{u}\right| _{bare}=-\left[ \varepsilon +\frac{5}{2}\bar{\zeta}_{\rho }-%
\bar{\zeta}_{\sigma }+\alpha \frac{v}{1+v}\left( 2-\bar{\zeta}_{\rho
}\right) \right] \bar{u},  \nonumber \\
\beta _{\bar{w}}\left( \bar{u},\bar{w},v\right) &=&\left. \mu \partial _{\mu
}\bar{w}\right| _{bare}=\left[ \bar{\zeta}_{\rho }-\bar{\zeta}_{\sigma
}+\alpha \frac{v}{1+v}\left( 2-\bar{\zeta}_{\rho }\right) \right] \bar{w}, 
\nonumber \\
\beta _{v}\left( \bar{u},\bar{w},v\right) &=&\left. \mu \partial _{\mu
}v\right| _{bare}=-\Biggl[ \left( 1+v\right) \bar{\zeta}_{\sigma }+\alpha
\left( 2-\bar{\zeta}_{\rho }\right) \Biggr] v,  \label{HBF}
\end{eqnarray}
and seek their fixed points $\left( \bar{u}^{*},\bar{w}^{*},v^{*}\right) $.
These fall into two groups.

The {\em first} group is characterized by $v^{*}=0$: In this case, Eqns.~(%
\ref{HBF}) reduce to the fixed point equations (\ref{SRBF}) for the SR
theory, so that we recover the equivalents of the familiar SR fixed points,
namely (i) the FDT satisfying $\left( 8\varepsilon /3+O\left( \varepsilon
^{2}\right) ,1,0\right) $, (ii) the FDT violating $\left( 16\varepsilon
+O\left( \varepsilon ^{2}\right) ,0,0\right) $, (iii) the fixed point $%
(16\varepsilon /3+O\left( \varepsilon ^{2}\right) ,1/3+O\left( \varepsilon
\right) ,0)$ on the separatrix between (i) and (ii), and finally the
Gaussian fixed line (iv) $\left( 0,\bar{w},0\right) $. In contrast to the SR
case, however, their stability needs to be investigated in the {\em three}%
-dimensional space spanned by $\left( \bar{u},\bar{w},v\right) $, controlled
by the $3\times 3$ matrix ${\Bbb M}\equiv (\partial _{i}\beta _{j})^{*}$, $%
i,j=\bar{u},\bar{w},v$. For $v^{*}=0$, we find $\partial _{\bar{u}}\beta
_{v}=\partial _{\bar{w}}\beta _{v}=0$ at all of these fixed points, so that
only the upper left $2\times 2$ corner of ${\Bbb M}$ (which is just ${\Bbb M}%
^{SR}$), and the bottom right element, $\partial _{v}\beta _{v}=-\bar{\zeta}%
_{\sigma }-\alpha \left( 2-\bar{\zeta}_{\rho }\right) $, are relevant. The
eigenvalues of ${\Bbb M}^{SR}$ were already computed in Section (3.1), so
that only fixed points (i) and (ii) remain as candidates for global
stability if $\varepsilon >0$. For fixed point (i), $\bar{\zeta}_{\rho }^{*}=%
\bar{\zeta}_{\sigma }^{*}=-2\varepsilon /3$ to all orders (cf. Section 3.1)
so that $\partial _{v}\beta _{v}$ is positive for $\alpha <\alpha _{1}\equiv
\varepsilon /\left( 3+\varepsilon \right) $. Thus, $\alpha _{1}$ demarcates
the stability boundary of the FDT satisfying SR fixed point (i). Considering
fixed point (ii), we have $\bar{\zeta}_{\rho }^{*}=-\varepsilon +O\left(
\varepsilon ^{2}\right) $ and $\bar{\zeta}_{\sigma }^{*}=-3\varepsilon
/2+O\left( \varepsilon ^{2}\right) $, giving positive values for $\partial
_{v}\beta _{v}$ provided $\alpha <\alpha _{2}\equiv 3\varepsilon /4+O\left(
\varepsilon ^{2}\right) $. So, the FDT violating SR fixed point (ii) remains
stable for a larger region of $\alpha $ than its FDT-satisfying partner. We
remark that, in contrast to $\alpha _{1}$, $\alpha _{2}$ is known only
perturbatively. Finally, the fixed line $\left( 0,\bar{w},0\right) $ is
stable provided both $\varepsilon $ and $\alpha $ are negative.

Returning to $\varepsilon >0$, it is clear that another fixed point must
become stable beyond $\alpha _{2}$. By necessity, this can only be a member
of the {\em second} group, having $v^{\ast }\neq 0$. For all of these, the $%
\left[ \ldots \right] $ bracket in the last line of Eqns.~(\ref{HBF})
vanishes, so that the first two equations simplify to 
\begin{eqnarray}
\beta _{\bar{u}}\left( \bar{u},\bar{w},v^{\ast }\right) &=&-\left[ \left(
\varepsilon +2\alpha \right) +\left( \frac{5}{2}-\alpha \right) \bar{\zeta}%
_{\rho }\right] \bar{u},  \nonumber \\
\beta _{\bar{w}}\left( \bar{u},\bar{w},v^{\ast }\right) &=&\Biggl[2\alpha
+\left( 1-\alpha \right) \bar{\zeta}_{\rho }\Biggr]\bar{w}.  \label{HBFS}
\end{eqnarray}
One should note that, with $\bar{\varepsilon}=\varepsilon +2\alpha $, $\beta
_{\bar{u}}$ is precisely the $\beta $-function of the LR theory, Eqn.~(\ref
{LRBF}). Thus, we identify the second group as the hybrid partners of the LR
fixed points. The equation for $\beta _{\bar{u}}$ has a trivial solution $%
\bar{u}^{\ast }=0$ and a nontrivial one, with $\bar{u}^{\ast }\neq 0$.
Seeking the corresponding values for $\bar{w}^{\ast }$, we obtain a
``Gaussian'' fixed point (v) $\left( 0,0,\infty \right) $ and a nontrivial
one (vi) with $\bar{u}^{\ast }=32\left( \varepsilon +2\alpha \right) /5$ $%
+O\left( \varepsilon ^{2}\right) $, $\bar{w}^{\ast }=0$ and $v^{\ast
}=\left( 4\alpha -3\varepsilon \right) /\left[ 3\left( \varepsilon +2\alpha
\right) \right] +2\alpha /3+O\left( \varepsilon ^{2}\right) $. Note that we
have expanded $\bar{u}^{\ast }$ in both $\varepsilon $ and $\alpha $,
keeping only terms to first order in either, since both are assumed to be of
the same order. $v^{\ast }$ is $O(1)$ which should not be disturbing since $%
v $ is not a perturbative coupling. Since $\bar{u}^{\ast }\neq 0$ here, we
find $\bar{\zeta}_{\rho }^{\ast }=-2\left( \varepsilon +2\alpha \right)
/\left( 5-2\alpha \right) $ to all orders.

Turning to the stability of these fixed points, it is straightforward to
determine that the Gaussian line is stable for $d>2\left( 1+\alpha \right) $%
, i.e., above the upper critical dimension of the LR theory. For $d<2\left(
1+\alpha \right) $, we find a more complex situation: for $\alpha <\alpha
_{1}$, fixed point (vi) has two unstable directions. One of these becomes
stable above $\alpha _{1}$. For $\alpha >\alpha _{2}$, this fixed point is
globally stable.

It is natural, of course, to seek fixed points where neither $\bar{u}^{*}$
nor $\bar{w}^{*}$ vanish. This attempt gives us a pair of equalities, valid
to all orders: 
\begin{eqnarray}
0 &=&\left( \varepsilon +2\alpha \right) +\left( \frac{5}{2}-\alpha \right) 
\bar{\zeta}_{\rho }^{*},  \nonumber \\
0 &=&2\alpha +(1-\alpha )\bar{\zeta}_{\rho }^{*},  \label{FL}
\end{eqnarray}
which do {\em not} result in a unique equation for such a fixed point.
Instead, they select a specific value of $\alpha $, namely $\alpha =\alpha
_{1}=\varepsilon /(3+\varepsilon )$, where a fixed {\em line} (vii) exists,
parameterized by $\bar{w}$: $\left( 32\varepsilon /\left[ 3\left( 3\bar{w}%
+1\right) \right] ,\bar{w},\left( 3\bar{w}-1\right) \left( 1-\bar{w}\right)
/\left( 3\bar{w}^{2}+2\bar{w}+3\right) \right) $. This line mediates the
stability loss of fixed point (i).

The crossover scenario between the SR and the true LR theory can now be
summarized. Let us fix $d$ just below $2$ and increase $\alpha $ starting
from zero. For sufficiently small $0\leq \alpha <\alpha _{1}$, the RG flow
is dominated by the SR behavior. The SR fixed points (i) and (ii) are
stable, within their respective domains of attraction. The separatrix forms
a surface which cuts the $v=0$ plane at $\bar{w}=1/3+O\left( \varepsilon
\right) $ and then bends over to larger values of $\bar{w}$ as $v$
increases. The unstable LR fixed point (vi) is found on the separatrix, in
the unphysical region $v<0$. When $\alpha $ reaches $\alpha _{1}$, the
correlated noise begins to make its presence felt. Specifically, for $\alpha
_{1}<\alpha <\alpha _{2}$, the FDT violating fixed point (ii) is the only
globally stable fixed point. The FDT restoring fixed point (i), while still
stable {\em within} the $v=0$ plane, has become unstable to small
perturbations {\em out} of that plane. Thus, a flow line starting near (i),
with a small $v>0$ component, will first flow out into the half-space $v>0$
and then bend {\em back} towards $v=0$, flowing into the FDT violating fixed
point (ii). The {\em sign} of $v$ remains invariant under the flow. The LR
fixed point (vi) is still unphysical, but has moved closer to the $v=0$
plane. Finally, at $\alpha =$ $\alpha _{2}$, (vi) merges with (ii) and moves
out into the positive $v$ region as $\alpha $ increases beyond $\alpha _{2}$%
. The LR fixed point (vi) is now the only stable one, and the global RG flow
is dominated by the LR theory. A different view of the SR-LR crossover is
presented in Fig. 1 which shows the location of the stability boundaries as
functions of $\alpha $ and $d$.

\vfill

\begin{figure}[tbph]
\vspace*{1.0cm}
\hspace*{2.0cm}
\epsfxsize=4in \epsfbox{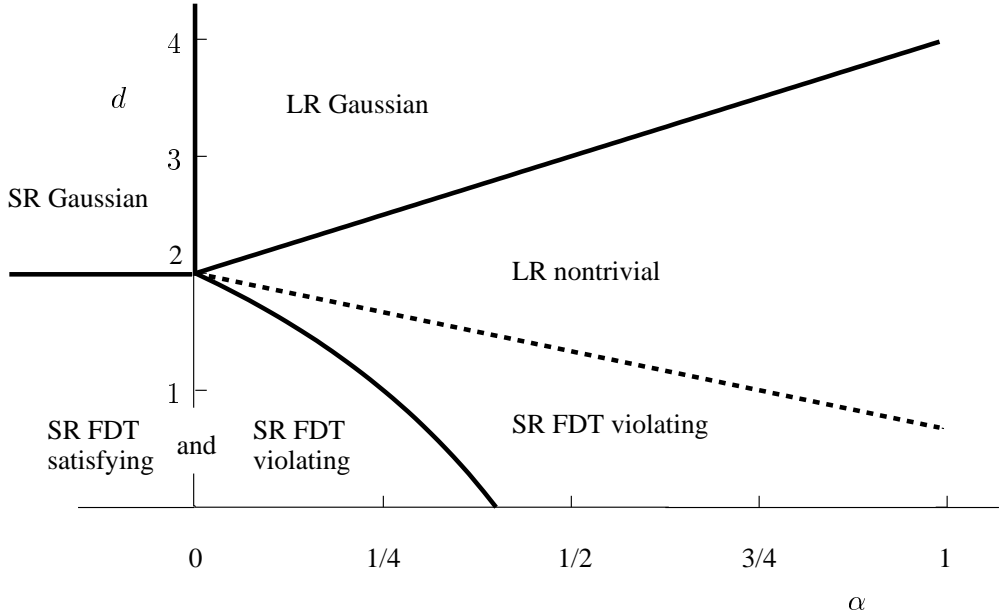}
\vspace*{0.5cm}
\caption{Stability boundaries of different fixed points as
functions of $d$ and $\alpha $. The heavy solid lines denote boundaries
whose
location is known to all orders in perturbation theory.}
\end{figure}

\vfill
 
It is interesting to note the two different scenarios which control the
stability loss of fixed points here. Fixed points (ii) and (vi) exchange
their stability by merging, similar to the stability exchange between the
Gaussian and the Wilson-Fisher \cite{WF} fixed points in $\phi ^{4}$-theory
when $d$ drops below $4$. In contrast, fixed point (i) loses its stability,
at $\alpha =\alpha _{1}$, by generating a {\em fixed line}. This fixed line
has two key properties: first, it connects fixed points (i), (iii) and (vi);
second, it lies on the surface forming the {\em separatrix} between fixed
points (i) and (ii). This surface bends back towards the $v=0$ plane, as $%
\alpha $ increases, to the extent that it {\em reaches} fixed point (i) at $%
\alpha =\alpha _{1}$. Thus, fixed point (i) loses one of its three stable
directions when being absorbed by the separatrix; moreover, this direction
is now spanned by the fixed line: this allows the flow to change sign when $%
\alpha $ crosses $\alpha _{1}$.

The crossover which we observe in the flow diagrams is also reflected by the
exponents. As an example, we consider the scaling form of the LR structure
factor, Eqn. (\ref{LRS}). We recall that its scaling behavior is determined
by the exponent $\Delta _{LR}=\bar{\varepsilon}/\left( 5-2\alpha \right) $.
The hybrid fixed point (vi), on the other hand, generates an exponent $%
\Delta \equiv -\bar{\zeta}_{\rho }^{*}/2=\left( \varepsilon +2\alpha \right)
/\left( 5-2\alpha \right) $. These two exponents clearly match, since $\bar{%
\varepsilon}=\varepsilon +2\alpha $. As $\alpha $ decreases, we reach the
stability boundary $\alpha _{2}=3\varepsilon /4+O(\varepsilon ^{2})$, where
(vi) and (ii) exchange stability. Here, the exponent associated with (ii) is 
$\Delta _{SR}=-\zeta _{\rho }^{SR*}/2=\varepsilon /2+O\left( \varepsilon
^{2}\right) $, to be compared with the corresponding value near (vi), $%
\Delta =(\varepsilon +2\alpha _{2})/(5-2\alpha _{2})$. In this way,
continuity is ensured. However, a discontinuous change of exponents may
occur as $\alpha $ decreases below $\alpha _{1}$: above $\alpha _{1}$, all
positive $\left( \bar{u},\bar{w},v\right) $ fall into the domain of
attraction of fixed point (ii), with $\Delta _{SR}=\varepsilon /2+O\left(
\varepsilon ^{2}\right) $; in contrast, below $\alpha _{1}$, some of these
will be attracted towards fixed point (i) where $\Delta _{SR}=\varepsilon /3$%
. For these theories, the strong anisotropy exponent $\Delta $ will change 
{\em discontinuously} upon crossing the stability boundary between fixed
points (i) and (ii). Note, however, that $\Delta _{SR}=\varepsilon /3$
coincides with $\Delta _{LR}=\bar{\varepsilon}/\left( 5-2\alpha \right) $ at
the line $\alpha =\alpha _{1}=\varepsilon /\left( 3+\varepsilon \right) $ to
all orders in $\varepsilon $. Nevertheless, even though the exponents may
undergo discontinuities, both the hybrid and the SR structure factors scale
according to Eqn.~(\ref{SRS}). Thus, the scaling {\em forms} remain
unchanged.

\section{Conclusions}

Using field theoretic methods, we have analyzed the RG flow for a model of
biased diffusion subject to a noise term, parameterized by its momentum
dependence, $q_{\Vert }^{2(1-\alpha )}$, with spatially long-ranged
correlations. One limit, $\alpha =0$, corresponds to a ``short-range'' model
with purely local, conserved noise, and the other, $\alpha =1$, models
biased diffusion with a non-conserved noise. The crossover from $\alpha <1$
to $\alpha =1$ presents no difficulties; however, the opposite limit, $%
\alpha \rightarrow 0$, is quite subtle. Here, the full crossover is observed
only if we interpose a hybrid model with $\alpha =O\left( \varepsilon
\right) $, between the SR ($\alpha =0$) and the LR ($\alpha $ finite)
theories. This hybrid contains the key elements of both, SR and LR, Langevin
equations. It possesses a number of fixed points, including the equivalents
of the SR and the LR models, so that the crossover can be understood in
terms of a stability exchange between different fixed points. Clearly,
considering $\alpha $ on the scale of $\varepsilon $ enables us to resolve
this ``fine structure''. Just below two dimensions, the scenario is
following: for $\alpha $ below a lower stability limit $\alpha _{1}$, the
theory is controlled by two stable SR fixed points, one FDT\ restoring and
the other FDT violating, each with its own basin of attraction. Stated
differently, we are in a ``weak noise'' regime, for $d<d_{LR}\equiv
2-3\alpha /\left( 1-\alpha \right) $, where the long-ranged noise does not
significantly modify the universal behavior of the system. As $\alpha $
increases beyond $\alpha _{1}$ but remains below an upper stability limit $%
\alpha _{2}$, the SR FDT restoring fixed point becomes unstable, leaving the
SR FDT violating fixed point in control of the flow. This stability exchange
is mediated by a fixed point line. Finally, above $\alpha _{2}$, the SR\ FDT
violating fixed point also destabilizes and the nontrivial LR fixed point
becomes globally stable. We note, in conclusion, that a similar mechanism,
namely stability exchange through a fixed point line, has previously been
observed in the Sine-Gordon model: there, two fixed points are stable below $%
d=2$, corresponding to the high- and low-temperature phases, respectively.
As $d$ increases beyond $2$, the high-temperature fixed point loses its
stability, via a similar fixed point line \cite{BN}.

Unfortunately, most of the nontrivial crossover phenomena discussed here are
confined to dimensions $1<d<$ $2$. Above $d=2$, the long-range noise
dominates, either via its nontrivial or its Gaussian fixed point. In one
dimension, on the other hand, there is no transverse subspace so that the
SR\ FDT violating fixed point cannot be accessed. Consequently, the scenario
described above must change significantly. It is conceivable that only the
lower stability limit might survive here or that the two stability limits
merge, leaving us with a stability exchange between the FDT restoring SR and
the LR fixed point. Since, at $\alpha =\alpha _{1}$, $\Delta _{SR}=\Delta
_{LR}=\varepsilon /3$ to all orders, this may be a reasonable conjecture.
Moreover, in $d=1$, we have $\alpha _{1}=1/4$ and $\Delta =1/3$. These
values agree with the corresponding results \cite{MHKZ,FJT} for the
one-dimensional KPZ equation with correlated noise where only one stability
limit is observed.

Nevertheless, our analysis plays the role of a pilot study for a number of
other interesting problems. Clearly, one might consider an interacting
theory subject to an external bias \cite{critDDS} and a long-range noise
term \cite{JSlong}. Here, $d_{c}=5$, so that physical dimensions are more
accessible, and comparisons with Monte Carlo simulations can be made. Other
questions of interest concern noise terms with different spatial
correlations, or nontrivial correlations in time. For equilibrium systems,
the existence of an underlying Hamiltonian ensures that these correlations
have no effect on static properties. For non-equilibrium steady states,
however, the static behavior is generically inseparable from the dynamics.
Studies of anomalous noise correlations in non-equilibrium systems may help
to unravel the nature of this coupling.

\bigskip

\acknowledgments

We wish to thank R.\ Bausch and R.K.P.\ Zia for valuable discussions. BS
also gratefully acknowledges the kind hospitality of the Institut f\"{u}r
Theoretische Physik III of the Heinrich-Heine-University of D\"{u}sseldorf
where some of this work was performed. This research is partially supported
by SFB 237 (``Unordnung und gro\ss e Fluktuationen'') of the Deutsche
Forschungsgemeinschaft and the US National Science Foundation through the
Division of Materials Research.


\end{document}